\documentclass[12pt]{article}
\usepackage{amssymb}
\usepackage{amsmath, amsthm}
\newcommand{\be}{\begin{equation}}
\newcommand{\ee}{\end{equation}}
\begin{document}
\def\theequation{\arabic{section}.\arabic{equation}}
\begin{titlepage}
\title{Extended quintessence, inflation, and stable de Sitter 
spaces}
\author{Valerio Faraoni and Michael N. Jensen\\ \\
{\small \it Physics Department, Bishop's University}\\
{\small \it Lennoxville, Qu\`{e}bec, Canada J1M~1Z7}
}
\date{} \maketitle
\thispagestyle{empty}
\vspace*{1truecm}
\begin{abstract}
A new gauge-invariant criterion for stability against 
inhomogeneous perturbations of de Sitter space is applied to 
scenarios of dark energy and inflation in scalar-tensor gravity. 
The results extend previous studies.
\end{abstract} \vspace*{1truecm}
\begin{center} {\bf PACS:}   98.80.-k, 04.90.+e, 04.50.+h
\end{center}
\begin{center}  {\bf Keywords:} dark energy, extended 
quintessence, inflation, de Sitter 
space
\end{center}
\setcounter{page}{1}
\end{titlepage}

\def\theequation{\arabic{section}.\arabic{equation}}


\section{Introduction}
\setcounter{equation}{0}
\setcounter{page}{2}

Recent studies of type Ia Supernovae \cite{SN} have demonstrated 
that the 
expansion of the universe is accelerated. In the context 
of Einstein gravity this can only be due to an exotic form of 
energy called dark energy or quintessence with negative  
pressure, which dominates the cosmic dynamics 
\cite{quintessence}.  The best fit of 
the 
observational data from cosmic microwave background 
experiments \cite{CMB} fixes the energy density of dark energy 
to be $\Omega_{de} = 0.7$ in units of the critical density, with 
matter accounting for $\Omega_m \simeq 0.3$, mostly composed of 
dark matter. The best fit of the observational data also  
favours a time-varying form of dark energy \cite{w} and there is 
marginal evidence for a negative equation 
of state parameter $w \equiv P/\rho < -1$, where $P$ and 
$\rho$ are the energy density and pressure of the dark energy, 
respectively \cite{w}.

Most models of dark energy are based on scalar fields, which 
are natural candidates from the theoretical point of view 
because it is well known from the inflationary theory of the 
primordial universe \cite{LiddleLyth} that a scalar field $\phi$ 
slowly rolling in 
a nearly flat section of its self-interaction potential 
$V \left( \phi \right) $ can fuel a period of accelerated 
expansion of the 
universe. However, a single canonical scalar field in Einstein 
gravity cannot explain the range of values $w < -1$ of the 
equation  of state parameter preferred by the observations 
\cite{Boisseauetal,superquintessence}.  In 
fact, a canonical scalar field $\phi$ in general relativity has 
energy density and pressure
\be \label{I1}
\rho = \frac{\dot{\phi}^2}{2} +V ( \phi ) \;, \;\;\;\;\;\;\;\;
 P= \frac{\dot{\phi}^2}{2} - V ( \phi ) \;,
\ee
where an overdot denotes differentiation with respect to the 
comoving time $t$. In the spatially flat 
Friedmann-Lemaitre-Robertson-Walker (FLRW) metric
\be \label{I2}
ds^2 = - dt^2 + a^2 (t) \left( dx^2 + dy^2 + dz^2 \right) 
\ee
in comoving coordinates $\left( t,x,y,z \right) $, the 
Einstein-Friedmann equation
\be \label{I3}
\dot{H} = -\frac{\kappa}{2} (\rho + P)
\ee
(where $\kappa \equiv 8\pi \, G$ and $H \equiv \dot{a} / a $ is 
the Hubble parameter) implies that $w < -1$ is equivalent to 
$\dot{H} > 0$, a regime called {\em superacceleration} as 
opposed to acceleration $ \ddot{a}/a= \dot{H} + H^2 > 0$. Then, 
eq.~(\ref{I3}) yields 
$\dot{H} = -\kappa\, \dot{\phi}^2/2  \leq 0$, with the upper 
bound of the inequality being attained by de Sitter spaces with 
$H = $const.  A regime with $\dot{H} > 0$ (or $w < -1$) 
requires an extension of the theory \cite{Boisseauetal}.  One 
possibility  that has 
received much attention in the literature 
\cite{Caldwell,BigRip,phantom} is that 
of phantom fields, i.e., the scalar $\phi$ is formally allowed 
to have a negative kinetic energy and  
$\dot{\phi}^2/2$ is replaced by $ -\dot{\phi}^2/2 $ in 
eq.~(\ref{I3}). This modification allows for dynamics capable of 
breaking 
the $w < -1$ barrier.  Motivations for phantom fields 
arise  in string/M  theory and in supergravity 
\cite{phantommotivations}.

Another possibility \cite{STquintessence} is to allow 
the scalar $\phi$  to couple nonminimally to the Ricci curvature 
through the action
\be\label{I4}
S = \int d^4x\, \sqrt{-g} \left[ \left( \frac{1}{\kappa} - \xi 
\, \phi^2 
\right) \, \frac{R}{2} -\frac{1}{2}\, 
g^{ab} \nabla_a \, \phi \, \nabla_b \,\phi -V( \phi) \right] \;.
\ee
Both phantom and non-minimally coupled scalar field models can 
be seen as scalar-tensor theories of gravity. Scalar-tensor 
models of dark energy (called  {\em extended quintessence}) were 
considered even before the  observational evidence for 
superacceleration was reported 
\cite{STquintessence} (see Refs.~\cite{mybook,FujiiMaeda} for 
reviews of cosmology in scalar-tensor gravity).  Perhaps the 
most striking consequence of superacceleration is that the 
universe 
may end its existence in a Big Rip singularity at a finite time 
in the future \cite{BigRip}. Whether a Big Rip is unavoidable 
or another fate awaits the late universe is determined by the 
presence and size of the attraction basins of  Big Rip solutions 
and/or other late time attractors in the phase space 
of the dark energy model. In certain models the universe 
eventually exits the superacceleration regime and converges to 
a de Sitter attractor, expanding for an 
infinite time \cite{deSitterattractors}. In other models the 
Big Rip is the only late time attractor 
(e.g., \cite{FaraoniBigRip,Faraoniphantom}).

In this paper we study the stability of de Sitter solutions 
(when they exist) in extended quintessence.  The  
generalized gravity action 
\be\label{I5}
S = \int d^4 x \, \sqrt{-g} \left[ \frac{f \left( \phi, R 
\right)}{2}  -\frac{\omega( \phi )}{2}\, g^{ab} 
\nabla_a \, \phi \, \nabla_b \,\phi -V( \phi) \right] 
\ee
contains scalar-tensor gravity as the special case in 
which the function $ f \left( \phi, R \right)$ is linear in $R$.  
The field 
equations in the FLRW metric (\ref{I2}) are
\begin{eqnarray} 
&& H^2  = \frac{1}{3F} \left( \frac{\omega}{2} \, \dot{\phi}^2 
+\frac{RF}{2} -\frac{f}{2} +V -3H\dot{F} 
\right)  \;, \label{I6} \\
&& \nonumber \\
&& \dot{H} =   - \, \frac{1}{2F} \left( \omega \dot{\phi}^2 + 
\ddot{F}  -H\dot{F} \right) \;, \label{I7}\\
&& \nonumber \\
&& \ddot{\phi } +3 H \dot{\phi} +\frac{1}{2\omega} \left(
 \frac{d\omega}{d\phi} \,  \dot{\phi}^2 - \frac{\partial 
f}{\partial \phi} +2\, \frac{dV}{d\phi}  
\right) =0 \;, \label{I8}
\end{eqnarray}
where $ F \equiv \partial f /\partial R $.
The Hubble parameter $H$ is a cosmological observable and the 
scalar $\phi$ is a natural dynamical variable to consider, hence 
we take $H$ and  $\phi$ as physical variables. Then  
the only equilibrium points of the dynamical system 
(\ref{I6})-(\ref{I8}) are de Sitter spaces with constant scalar 
field $\left( H_0, \phi_0 \right)$ and they exist subject to the 
two conditions
\begin{eqnarray} 
&& 6H_0^2 \, F_0 - f_0+2V_0 =0 \;, \label{I9} \\
&& \nonumber \\
&& {f_0}'- 2{V_0}' =0 \;, \label{I10} 
\end{eqnarray}
where 
$F_0 \equiv F \left( H_0, \phi_0 \right)$, $f_0 \equiv f \left( 
\phi_0, R_0 \right)$, $V_0 \equiv V  \left( \phi_0 \right)$, 
${V_0}' \equiv \left. \frac{dV}{d\phi}  \right|_{ \phi_0 }$, 
${f_0}' \equiv \left. \frac{\partial f}{ \partial \phi} 
\right|_{\left( \phi_0, R_0 \right)}$, and $R_0 = 12 H_0^2$.
Equations (\ref{I9}) and (\ref{I10}) are straightforward to 
derive from 
eqs.~(\ref{I6})-(\ref{I8}) and appeared in the literature 
before for various special choices of the function $ 
f\left( \phi, R  \right) $ 
\cite{BarrowOttewill}-\cite{Ohta}. 
There are only two conditions 
because only two equations in this set are independent.    
The stability of these de Sitter spaces is usually assessed 
with respect to  homogeneous (time-dependent only) 
perturbations; in Ref.~\cite{deSitterPRD} we studied  
stability with respect to inhomogeneous (space- and 
time-dependent) perturbations, to linear order.  The difficulty 
with inhomogeneous perturbations is related to their 
gauge-dependence and  to deal with this problem we used the 
covariant and gauge-invariant formalism of Bardeen, Ellis, 
Bruni, Hwang and Vishniac \cite{Bardeen}, in the formulation 
given by Hwang and 
Hwang and Noh for generalized gravity \cite{Hwang}. We found  
that, when $\phi_0 \neq 0$, the de Sitter fixed point $\left( 
H_0, \phi_0 \right) = \left( \sqrt{\frac{f_0 - 2V_0}{6F_0}}, 
\phi_0 \right)$ is stable if and only if
\be \label{I11}
\frac{  \left( \frac{f_0''}{2} -  V_0'' +\frac{6 \left( 
H_0 f_{\phi R} \right)^2}{F_0} \, 
 \right)}{\omega_0 \, \left( 1+\frac{ 3f_{\phi R}^2}{2\omega_0 
F_0} \right)} \leq 0 
\ee
(see Refs.~\cite{BarrowOttewill,NVA,Zerbini,CliftonBarrow} for 
similar conditions in generalized gravity).  A different formula 
applies if $\phi_0 = 0$ but we will not be 
concerned with this case here, while contracting 
de Sitter fixed points with $H_0 = -\sqrt{\frac{f_0 - 
2V_0}{6F_0}}$ are always unstable \cite{deSitterPRD}. 
We refer the reader to the Appendix for a sketch of the 
derivation of the stability criterion (\ref{I11}), and to 
Ref.~\cite{deSitterPRD} for  a detailed discussion.

Our goal is to apply the stability condition (\ref{I11}) to 
extended quintessence scenarios that appeared recently in the 
literature and to scalar-tensor inflationary scenarios of the 
early universe \cite{extendedhyperextended}.

\section{Non-minimally coupled quintessence}
\setcounter{equation}{0}

We begin with the extended quintessence model of Carvalho \& 
Saa  \cite{CarvalhoSaa} based on a non-minimally coupled scalar 
field  and described by
\be \label{I12}
f \left( \phi, R \right )= (1-\xi\phi^2)R \;, \;\;\;\;\; \omega 
\equiv 1\;, \;\;\;\;\; V( \phi)=A \, \mbox{e}^{-\sigma \, \phi} \;.
\ee
where $\xi < 0$ is a dimensionless coupling constant, $ A $ and 
$\sigma$ are positive constants, and units are used in which 
$ \kappa= 8\pi \, G = 1$.  Equations (\ref{I9}) and 
(\ref{I10}) for the 
existence of de Sitter fixed points $\left( H_0, \phi_0 \right)$ 
yield
\be \label{I13}
H_0^2 = \frac{V_0}{3(1 - \xi \, \phi_0^2)} = 
\frac{\sigma \, A \, \mbox{e}^{-\sigma \, \phi}}{12\xi \, 
\phi_0} \;, \;\;\;\;\;\;\;\;
 \phi_0^{(\pm)} = -\, \frac{2}{\sigma}\left( 1 \pm \sqrt{1 + 
\frac{\sigma^2}{4\xi} } \, \right) \;,
\ee
which correspond to eqs.~(14) and (13) of 
Ref.~\cite{CarvalhoSaa}, respectively. These authors consider 
the range of values of the coupling constant $\xi < 
-\sigma^2/4 $, and $\phi_0^{\pm} < 0$. The criterion for 
stability (\ref{I11}) in the theory (\ref{I12}) reduces to the 
one previously found in 
Ref.~\cite{FaraoniPLA} and can be written as
\be \label{I14}
V_0'' \geq \varphi ( x) \, \frac{V_0'}{\phi_0} \;, 
\,\;\;\;\;\;\;\;\;\; 
x= \xi \phi_0^2 \;, \;\;\;\;\;\;\;\; \varphi(x)=\frac{ 
1-3x}{1-x} \;,
\ee
if $\phi_0 \neq 0$.  It is convenient to use the variable $ y 
\equiv \frac{\sigma^2}{4 \left|\xi\right|}$ with $0 < y < 1$. In 
the case of the fixed point
\be \label{I15}
\left( H_0, \phi_0^{(-)} \right) = \left( 
\sqrt{ \frac{V_0}{ 3\left[ 1 - \xi (\phi_0^{(-)})^2 \right] }} ,
\, -\, \frac{2}{\sigma}\left( 1 + \sqrt{1 + 
\frac{\sigma^2}{4\xi} } \right) \right) \;,
\ee
the stability condition (\ref{I14}) simply reduces to 
\be
1-y+\sqrt{1-y} 
 \geq 0
\ee
which is always satisfied, hence this equilibrium 
point is stable against linear inhomogeneous perturbations. This 
conclusion complements  the work of Carvalho and Saa 
\cite{CarvalhoSaa} who consider only homogeneous perturbations 
but to arbitrary order by exhibiting a suitable  Ljapunov 
function.  On the one hand their approach provides information 
on the attraction  basin; on the other hand, our approach 
takes into account the more general 
inhomogeneous perturbations by tackling the gauge-dependence 
problems. Regarding the parameter space, Carvalho and Saa  
only prove stability analytically for $-5/6 < \xi < 
- \sigma^2/4 $, although they conjecture stability for a much 
larger range of values of this parameter on the basis of 
numerical integrations probing down to values $\xi < -100$.  Our 
approach instead establishes linear stability analytically for 
any $\xi < - \sigma^2/4 $. It is to be noted however that 
the values of $\sigma$ and $\xi$ are not arbitrary:  
they are subject to bounds placed by observational constraints 
on the variability  of the effective gravitational coupling 
$G_{eff}(\phi) = G\left( 1 - 8\pi \, G \, \xi \, \phi^2 
\right)^{-1} $ ~~\cite{Gconstraints}.

The second fixed point $\left(H_0, \phi_0^{(+)} \right)$ is 
found to be unstable against homogeneous perturbations in 
Ref.~\cite{CarvalhoSaa}; this is of course true also 
for inhomogeneous perturbations.  In fact the stability 
condition (\ref{I14}) reduces to $y(1-y) \leq 0$, which is never 
satisfied because $y$ only assumes values in the range $0 < y < 
1$.
     
The constraints on the variation of $G_{eff}$ mentioned 
above   \cite{Gconstraints} do not 
apply  to inflation but they are relevant to more recent epochs 
of the cosmic history (e.g., during galaxy formation) and to 
scenarios of dark energy in the present universe. 
Although most constraints depend on specific quintessence models 
or parametrizations of the variation of $G_{eff}$, the general 
feeling is that it is rather difficult to reconcile the 
available constraints on the variation of $G_{eff}$ with a 
successful extended quintessence scenario. While the latter 
requires low energy scales, high energy scales are generally 
required to implement the modest variations of $G_{eff}$ 
allowed. The realization of a successful extended quintessence 
scenario compatible with all the observational constraints and 
free of fine-tuning problems may in the end  reveal to be very 
problematic \cite{Gconstraints} and this issue deserves further 
investigation.

\section{Inflation with non-minimal coupling   }
\setcounter{equation}{0}

The approach presented here is suitable to study also
the stability of de Sitter-like inflation in
scalar-tensor gravity; in this context the presence of 
a de Sitter 
attractor justifies the use of  the slow-roll approximation to 
inflation in
the early universe \cite{LiddleLyth}.  We turn our
attention to the recent non-minimally coupled scalar
field inflationary models of
Refs.~\cite{KomatsuFutamase}-\cite{TsujikawaGumjudpai} (see also 
\cite{Perrottaetal}) which 
employ the power-law
potential $V(\phi) = C \phi^p$.  The conditions
(\ref{I9}) and (\ref{I10}) for the existence of de
Sitter fixed points $\left( H_0, \phi_0 \right)$ yield
\be \label{I16} 
H_0^2 = \frac{C\left( 2-p \right) \,
\phi_0^p}{6 \left( 1+\xi \, \phi_0^2 \right)} =
-\frac{C}{12} \, p^{p/2} \,\left( p-4 \right)^{1-p/2}
\,\, \xi^{-p/2} \;, \;\;\;\;\;\;\;\;\; \phi_0^2 =
\frac{p}{(p-4)\xi} \;. 
\ee 
These expressions correspond, e.g., to eqs.~(3.4) and (3.5)
of Ref.~\cite{TsujikawaGumjudpai}.  If $\xi < 0$ it
must be $0 \leq p < 4$ for the existence of these fixed
points while if $\xi > 0$ it must be $p \leq 0$ or $p >
4$.  The stability condition (\ref{I14}) yields,
assuming a positive effective gravitational coupling, i.e., $1
- \xi \, \phi_0^2 > 0$,
\be
\phi_0^2 \leq \frac{2-p}{(4-p)\xi} \;.
\ee
By using the expression of $\phi_0^2$ in (\ref{I16}) and 
remembering the condition for the existence of de Sitter space, 
one concludes that:

\begin{itemize}

\item If $\xi>0$ and $p\leq 0$ the fixed point $\left( H_0, 
\phi_0 \right)$ is {\em stable}.

\item If $\xi>0$ and $p > 4 $ this fixed point is {\em stable}.

\item If $ \xi>0$ and $0<p \leq 4$ this fixed point is {\em 
unstable}.

\item If $\xi < 0$ and $ 0 \leq p \leq  4 $ this fixed point is 
{\em unstable}.

\end{itemize}

Therefore, de Sitter attractors exist in phase space for $\xi>0$ 
and for $p\leq 0$ and $p>4$.  The requirement 
that the de Sitter point be an attractor does not constrain 
$\xi$ apart from its sign, but the spectrum of density 
perturbations generated during inflation does 
set constraints  on $\xi$ for various 
choices of the power $p$ \cite{NMCconstraints}.


\section{Induced gravity inflation}
\setcounter{equation}{0}

We now turn our attention to induced gravity 
\cite{IG}, which has been 
used as the context for both dark energy  and 
inflationary  models with power law potential 
\cite{CooperVenturi,IGinflation}.  
This theory is described by
\be \label{I17}
f \left( \phi, R \right )= \epsilon \phi^2 \, R \;, 
\;\;\;\;\;\;\;\;\;\;
 \omega \equiv 1\;, \;\;\;\;\;\;\;\; V( \phi)= C\, \phi^p \;.
\ee
Conformal coupling corresponds to $\epsilon=-1/6$. Here we 
restrict ourselves to considering positive gravitational 
coupling $\epsilon>0$.\footnote{For negative gravitational 
coupling $\epsilon<0$ de Sitter solutions with the quartic 
potential $V=\lambda\phi^4/4$ exist only if $\lambda<0$.} By 
using eqs. (\ref{I9}) and (\ref{I10}) one finds that
\be  \label{42}
H_0^2=\frac{\left( p-2\right)C \phi_0^{p-2}}{6\epsilon}
\ee
and it would appear that any power $p$ in the potential gives an 
admissible de Sitter space, but further comparison with 
eq.~(\ref{I9}), which assumes the form $ -3 \epsilon H_0^2 
+C\phi_0^p=0$ yields the result that de Sitter equilibrium 
points $\left( H_0, \phi_0 
\right)$ can only exist if $p = 4$. In this case, a mass term 
$ m^2\phi^2/2 $ can not be added to the the potential if one 
wants to find de Sitter solutions and use them as 
inflationary attractors.  For $V (\phi) =\lambda  \, \phi^4/4 $ 
the de 
Sitter fixed points are  given by\footnote{This expression 
matches, e.g., eq.~(7d) of Ref.~\cite{CooperVenturi}.}
\be\label{I18}
\left( H_0, \phi_0 \right) = \left(\pm \, \sqrt{\frac{\lambda 
\, \phi_0^2}{12\epsilon}}, \phi_0 \right) \;,
\ee
including the Minkowski space 
$\left( H_0, \phi_0  \right)= \left( 0, 0 
 \right)$.  Discarding the contracting de Sitter spaces with 
$H_0 < 0$ that are always unstable \cite{deSitterPRD}, the 
stability condition (\ref{I11}) reduces to $ 12\epsilon H_0^2 
-\lambda\phi_0^2 \leq 0$. By using eq.~(\ref{42}) it is seen 
that this stability condition is always satisfied for 
$\epsilon > 0$  and $\lambda > 0$ and the fixed point 
$\left(\sqrt{\frac{\lambda \, \phi_0^2}{2\epsilon}}, \phi_0 
\right)$ is stable, to first order, with respect to 
inhomogeneous perturbations.  To the best of our knowledge, only 
stability against homogeneous perturbations was considered in 
the literature.  

\section{Conclusions}
\setcounter{equation}{0}

The criterion~(\ref{I11}) for the stability of de Sitter space 
in scalar-tensor gravity allows one to quickly assess stability 
of de Sitter fixed points given the scalar-tensor Lagrangian of 
the theory, once conditions (\ref{I9}) and (\ref{I10}) establish 
their existence. The advantage of this stability condition is 
that it establishes stability with respect to the more general 
inhomogeneous perturbations and already incorporates a 
covariant and gauge-invariant treatment. By contrast, most 
studies of stability available in the literature are restricted 
to the less general homogeneous perturbations.  The main 
limitation of the stability criterion (\ref{I11}) is that its 
validity is restricted to linear perturbations. While this 
suffices for most situations, second or higher order 
corrections would be needed, e.g., to analyze the backreaction 
of long wavelength perturbations on the de Sitter metric.

The simplification introduced by (\ref{I11}) in the stability 
analysis of de Sitter space is made clear by the scenarios 
discussed in this paper in which we can quickly recover the 
presence of de Sitter attractors in one scenario of 
extended quintessence and in two scalar-tensor inflationary 
scenarios, and extend the results of previous studies. There are 
other scenarios in the literature for which one would 
like to be reassured in a gauge-independent way about the 
stability of a crucial de Sitter attractor with respect to  the 
more general  inhomogeneous perturbations.  These scenarios will 
be analyzed in future publications.

\section*{Acknowledgments}

This work was supported by the Natural Sciences and Engineering 
Research Council of Canada  (NSERC).

\vskip1truecm

\section*{Appendix: Derivation of the stability criterion 
(\ref{I11})}
\def\theequation{A.\arabic{equation}}
\setcounter{equation}{0}

Here  the stability criterion (\ref{I11}) is derived using the 
covariant and gauge-invariant formalism of Bardeen in 
the form 
developed by Ellis, Bruni Hwang, and Vishniac 
\cite{Bardeen}. A  version of this formalism directly applicable 
to scalar-tensor gravity was given in Refs.~\cite{Hwang}. The 
metric  perturbations are defined by the equations 
\begin{eqnarray}
g_{00} &= & -a^2 \left( 1+2AY \right) \;, \label{19}\\
&& \nonumber \\
g_{0i} & = & -a^2 \, B \, Y_i  \;, \label{20}\\
&& \nonumber \\
g_{ij} & =& a^2 \left[ \gamma_{ij}\left(  1+2 H_L \right) +2H_T 
\,  Y_{ij}  \right] \;,\label{21}
\end{eqnarray}
where the scalar harmonics satisfy  the eigenvalue problem
\be \label{22}
\bar{\nabla_i}
\bar{\nabla^i} \, Y =-k^2 \, Y \;,
\ee
$\gamma_{ij} $ are the components of the 
three-dimensional FLRW background metric, and $ \bar{\nabla_i}  
$ is  the covariant derivative of  $ \gamma_{ij}$. 
The defining relations for the vector and tensor 
harmonics $Y_i$ and $Y_{ij}$ are
\be \label{23}
Y_i= -\frac{1}{k} \, \bar{\nabla_i} Y \;, \;\;\;\;\;\;
Y_{ij}= \frac{1}{k^2} \, \bar{\nabla_i}\bar{\nabla_j} 
Y +\frac{1}{3} \, Y \, \gamma_{ij} \;.
\ee
The variables used are the Bardeen potentials 
$\Phi_H $ and $\Phi_A$   and the 
Ellis-Bruni  variable $\Delta \Phi $ given by \cite{Bardeen}
\be \label{25}
\Phi_H = H_L +\frac{H_T}{3} +\frac{ \dot{a} }{k} \left( 
B-\frac{a}{k} \, \dot{H}_T \right) \;, 
\ee

\be \label{26}
\Phi_A = A  +\frac{ \dot{a} }{k} \left( B-\frac{a}{k} \, \dot{H}_T \right)
+\frac{a}{k} \left[ \dot{B} -\frac{1}{k} \left( a \dot{H}_T \right)\dot{}  \right] \;, 
\ee

\be \label{27}
\Delta \phi = \delta \phi  +\frac{a}{k} \, \dot{\phi}  \left( B-\frac{a}{k} \, \dot{H}_T 
\right) 
\;.
\ee
The gauge-independent variables $\Delta F, \Delta 
f$, and $\Delta R$ are introduced by equations similar to 
eq.~(\ref{27}). 

By specifying that the background spacetime is a de Sitter one, 
the perturbations satisfy, to first order, the equations
\be \label{38}
\Delta \phi =\delta \phi \;, \;\;\;\; \Delta R=\delta R\; , 
\;\;\; \Delta F=\delta F \;, \;\;\;\; 
\Delta f =\delta f \;.
\ee
The first order equations ruling the evolution of these 
perturbations reduce to \cite{deSitterPRD}
\be  \label{42bis}
\Delta \ddot{\phi} + 3H_0  \Delta \dot{\phi} 
+ \left[ \frac{k^2}{a^2}-\,  \frac{1}{2\omega_0} \left( f_0''   
- 2 V_0'' \right) \right] 
\Delta \phi  =
 \frac{ f_{\phi R}}{2 \omega_0 } \,  \Delta R  \; ,
\ee

\be  \label{43}
\Delta \ddot{F} +3H_0 \, \Delta \dot{F} +\left( \frac{k^2}{a^2} - 4H_0^2 \right) \Delta F 
+\frac{F_0}{3} \, \Delta R =0 \;,
\ee

\be \label{44}
\ddot{H}_T +3H_0  \, \dot{H}_T +\frac{k^2}{a^2} \, H_T=0 \;,
\ee

\be \label{45}
-\dot{\Phi}_H+H_0 \Phi_A =\frac{1}{2} \left( \frac{\Delta \dot{F}}{F_0} -H_0 \, \frac{ \Delta 
F}{F_0} 
\right) \;,
\ee

\be  \label{46}
\Phi_H  = -  \frac{1}{2} \, \frac{ \Delta F}{F_0}   \; ,
\ee

\be   \label{47}
\Phi_A + \Phi_H =  - \frac{\Delta F }{F_0} \; ,
\ee

\be  \label{48}
\ddot{\Phi}_H + 3H_0 \dot{\Phi}_H  - H_0  \dot{\Phi}_A -3H_0^2 \Phi_A  
 =  - \frac{1}{2}\frac{  \Delta \ddot{F}}{F_0} - H_0 \, \frac{\Delta \dot{F}}{F_0} 
+ \frac{3H_0^2 }{2} \, \frac{\Delta F}{ F_0 }   \; ,
\ee
where $\Delta R$ is now given by
\be \label{49}
\Delta R=6 \left[ \ddot{\Phi}_H + 4H_0 \dot{\Phi}_H + 
\frac{2}{3} \frac{k^2}{a^2} \, \Phi_H 
-H_0 \dot{\Phi}_A + \left( \frac{k^2}{3a^2} -4H_0^2 \right) 
\Phi_A \right] \;,
\ee
and with the notations
\be
 f_{\phi R} \equiv \left.  \frac{ \partial^2 f}{ \partial \phi 
\partial R}  \right|_{ \left( \phi_0, R_0 \right)} \;, 
\;\;\;\;\;\;
f_0'' \equiv \left. \frac{\partial^2 f}{\partial \phi^2} 
\right|_{\left( \phi_0, R_0 \right)} \;.
\ee

By comparing eqs.~(\ref{46}) and (\ref{47}) one obtains
\be \label{50}
\Phi_H=\Phi_A =-\, \frac{\Delta F}{2F_0} \;.
\ee
Further  substitution in eq.~(\ref{49}) yields
\be \label{52}
\Delta R=6 \left[ \ddot{\Phi}_H + 3H_0 \dot{\Phi}_H + \left( 
\frac{k^2}{a^2}  -4H_0^2 \right) \Phi_H 
 \right] \;.
\ee

The stability with respect to tensor perturbations was studied 
in Ref.~\cite{deSitterPRD}, with the result that de Sitter space 
is always 
stable, to first order, with respect to these modes. The 
vorticity modes cannot be generated in scalar-tensor (or 
generalized) gravity theories when matter contributions are 
absent. There remain the scalar modes. We now restrict our 
attention to scalar-tensor gravity, for which $f \left( \phi, R 
\right)$ is linear in the Ricci curvature $R$. By using $ \frac{ 
\Delta F}{F_0} = \frac{ f_{\phi R}}{F_0} \, \Delta \phi $, 
eq.~(\ref{50}) gives (see Ref.~\cite{deSitterPRD} for further 
details)
\be \label{87}
\Phi_H= -\, \frac{1}{2} \, \frac{f_{\phi R}}{F_0} \, \Delta 
\phi 
\ee
while 
\be \label{88}
\Delta R= -\, \frac{3 f_{\phi R} }{ F_0 } \left[  \Delta 
\ddot{\phi} +3H_0 \, \Delta 
\dot{\phi} + \left( \frac{k^2}{a^2} -4H_0^2 \right) \Delta \phi 
\right] \;.
\ee
from eqs.~(\ref{52}) and (\ref{87}). This, and  
eq.~(\ref{42bis}), in turn yield
\be \label{89}
\Delta \ddot{\phi} + 3H_0 \, \Delta \dot{\phi} 
+  \left[ \frac{k^2}{a^2}-\,  \frac{  \left( \frac{f_0''}{2}   -  
V_0'' +\frac{6 f_{\phi R}^2}{F_0} \, 
H_0^2 \right)}{\omega_0 \, \left( 1+\frac{ 3f_{\phi R}^2}{2\omega_0 F_0} \right)}   \right] 
\Delta \phi  = 0  
\ee
when  $ 1+3f_{\phi R}^2/( 2 \omega_0 F_0) \neq 0$. When instead 
$ 1+3f_{\phi R}^2/( 2 \omega_0 F_0) = 0 $,   eq.~(\ref{42bis}) 
has either the trivial solution $\Delta \phi=0$ or $ 
f_0''-2V_0''=8\omega_0 \, H_0^2$.
 
The choice 
\be \label{90}
\Delta \phi =\epsilon \, \mbox{e}^{\gamma \, t} \;,
\ee
provides solutions of the asymptotic form of eq.~(\ref{89})  at 
late  times. Here $ \epsilon $ and  $\gamma $ are constants and  
$\gamma$  obeys the algebraic equation
\be \label{91}
\gamma^2 +3H_0 \, \gamma + c =0 
\ee
with
\be \label{91bis}
c = -\, \frac{  \left( \frac{f_0''}{2}   -  V_0'' +\frac{6 f_{\phi R}^2}{F_0} \, 
H_0^2 \right)}{\omega_0 \, \left( 1+\frac{ 3f_{\phi R}^2}{2\omega_0 F_0} \right)} \;.
\ee
The roots of of eq.~(\ref{91bis}) are 
\be 
\gamma_{\pm} =\frac{ -3H_0 \pm \sqrt{ 9H_0^2-4c}}{2} \;.
\ee
It is straightforward to see that  $Re\left( \gamma_{-} \right) 
<0$, while the sign of $Re\left( \gamma_{+} \right)$ 
depends on the sign of~$c$. If $ c \geq 0$ then $Re\left( 
\gamma_{+} \right) \leq 0 $ and we have  
stability. If  instead  $c < 0$ then $ Re\left( \gamma_{+} 
\right)  > 0 $ and de Sitter space  is 
unstable.  The  condition for the stability of de Sitter space 
in the scalar-tensor theory (\ref{I5}) with $f(\phi, R) $ 
linear in $R$ is therefore
\be\label{93}
\frac{   \left( \frac{f_0''}{2}   -  V_0'' +\frac{6 f_{\phi 
R}^2}{F_0} \,  H_0^2 \right)}{\omega_0 \, \left( 1+\frac{ 
3f_{\phi R}^2}{2\omega_0 F_0} \right)} \leq 0 \;.
\ee


\end{document}